\documentclass[prl,preprint,showpacs,superscriptaddress]{revtex4}

\usepackage{amsmath}
\usepackage{epsfig}
\usepackage{rotating}

\begin{document}

\title{Prediction of Giant Electro-actuation for Carbon Nanoscrolls}

\author{R. Rurali}
\affiliation{Laboratoire Collisions, Agr\'{e}gats, R\'{e}activit\'{e}, 
             IRSAMC, Universit\'{e} Paul Sabatier, 
             118 route de Narbonne, 31062 Toulouse cedex, 
             France}
\affiliation{Departament d'Enginyeria Electr\`{o}nica, 
             Universitat Aut\`{o}noma de Barcelona
             08193 Bellaterra, Spain}

\author{V.~R.~Coluci}
\author{D.~S.~Galv\~{a}o}
\affiliation{Instituto de F\'{i}sica "Gleb Wataghin", 
             Universidade Estadual de Campinas,
             C.P. 6165, 13083-970 Campinas SP, Brazil}

\date{\today}

\begin{abstract}
We study by first-principles calculations the electro-mechanical
response of carbon nanoscrolls. We show that although they  
present a very similar behavior to carbon nanotubes for what 
concerns the axial deformation sensitivity, they exhibit a radial 
response upon charge injection which is up to one order of magnitude 
larger. In association with their high stability, this behavior 
make them a natural choice for a new class of very efficient 
nano-actuators.
\end{abstract}


\maketitle

\begin{figure}[t]
\begin{center}
\epsfxsize=14cm
\epsffile{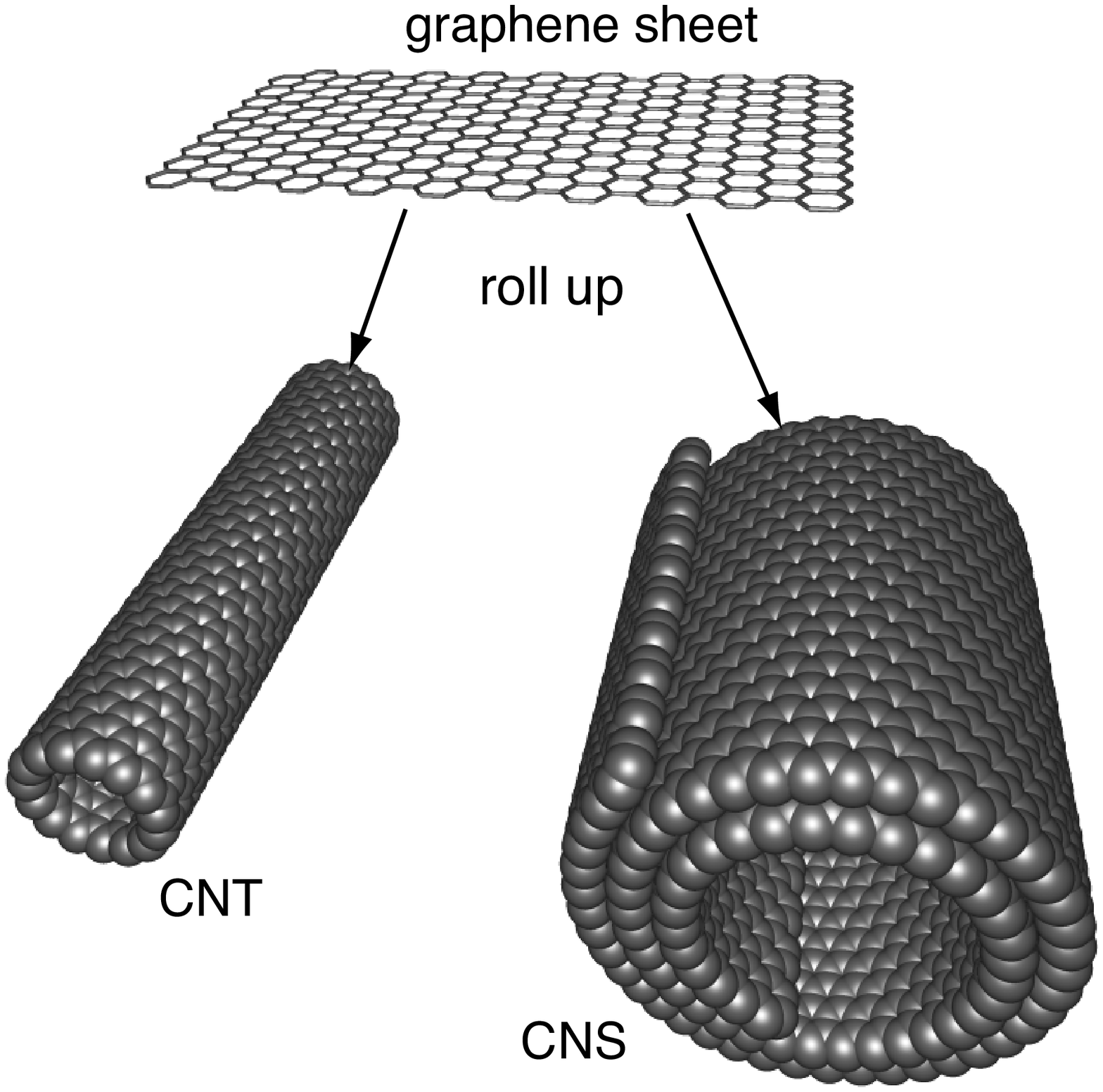}
\end{center}
\caption{Carbon nanotubes and scrolls can be topologically
         considered as cylindrical and papyrus-like structures,
         respectively, obtained from rolled up graphene layers.}
\label{fig:cnt_cns}
\end{figure}

With the advent of nanotechnology a great effort has been devoted
to the study of nanostructures, carbon nanotubes (CNTs)~\cite{ijima} 
being one of the most studied. In spite of more than two decades of 
intense research, the detailed mechanism of tube formation remains 
unclear~\cite{deheer}. It has been proposed that CNTs could be a 
subsequent state of papyrus-like carbon structures, generally named 
carbon nanoscrolls (CNSs)~\cite{bacon,zhou,ruland,shioyama,viculis,
khang,lauin}(Fig.~\ref{fig:cnt_cns}).

CNSs are remarkable structures sharing some of the rich mechanical
and electronic properties exhibited by CNTs and potentially presenting
new ones. They are known since the sixties from the pioneering work
of Bacon~\cite{bacon} who first reported the growth of scroll whiskers.
Surprisingly, very few studies~\cite{bacon,zhou,ruland,shioyama,viculis,
khang,lauin,ameclinckx,mordkovich,maniwa,tomanek,tomanek2,braga,pan} 
have been carried out for these systems. This can be explained in part by
the intrinsic experimental difficulties in synthesis, purification,
isolation, and characterization. However, after the recent advances
in the low-temperature synthesis of CNSs~\cite{shioyama,viculis,khang} 
there is a renewed interested in these materials.
Like CNTs, CNSs can be made of a single graphene sheet or by many of them.
However, in contrast to CNTs, the scroll diameter can vary easily
(expand or contract), thus they are extremely radially flexible. This
property can be explored for a variety of technological applications, 
such as chemical doping, hydrogen storage, electro-actuation
(mechanical deformation upon charge injection), etc..

The electro-mechanical response of CNTs has been investigated by
means of first-principles calculations by Verissimo-Alves
{\em et al.}~\cite{verissimo} and using an electron-lattice model
by Gartstein {\em et al.}~\cite{gartstein}. The electron actuation 
effects are predicted to occur, but of limited magnitude -~0.2-0.3\%~-
since the deformation of sp$^2$ carbon bond-lengths in close tubular 
structures as CNTs will require a significant amount of energy. 
An experimental demonstration of CNT based actuator has been reported
by Baughman {\em et al.}~\cite{baughman}. CNSs, on the other hand, 
are open structures and the radial expansion to accommodate the injected 
charges should be energetically more favorable.
In this case it has simply to overcome the van der Waals interlayer 
interactions, instead of deforming sp$^2$ carbon bond-lengths, thus 
producing a more significant electro-actuation behavior.
Recently, Braga et al.~\cite{braga} have used classical molecular dynamics
simulations to predict that CNSs should exhibit a significant radial expansion
upon charge injection. However, it is not possible to have a reliable 
quantitative estimation using classical methods because quantum effects 
are not included, as well as it is not possible to differentiate between 
electrons and hole injections, which is known to produce a different response 
for CNTs~\cite{verissimo,gartstein}. In order to properly address these 
issues the use of full quantum methods is necessary. In this work we report 
such study for some selected scroll models.

\begin{figure}[t]
\begin{center}
\epsfxsize=14cm
\epsffile{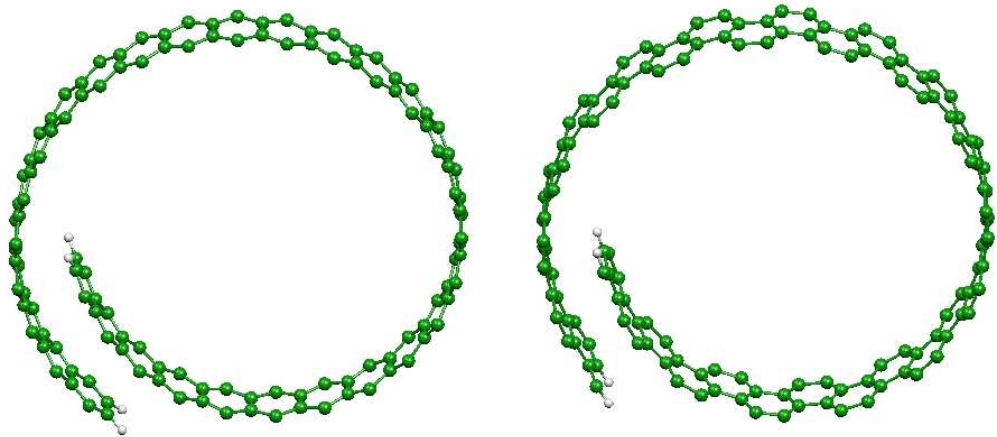}
\end{center}
\caption{(color online) Unit cells used in the calculations of
         (a) an armchair and (b) a zig-zag CNS. We have followed
         the nomenclature introduced by Braga {\em et al.}~\cite{braga},
         in turn derived from previous works on carbon
         nanoribbons~\cite{fujita,nakada}.}
\label{fig:unit_cells}
\end{figure}

We have carried out density-functional theory (DFT) calculations
in the framework of the local density approximation (LDA) with the 
{\sc Siesta} code~\cite{siesta}. We have used a double-$\zeta$ basis 
set plus polarization functions and norm-conserving pseudopotentials
of the Troullier-Martins type~\cite{troullier:martins}. We have 
considered two prototype structures in supercell geometry 
(see Fig.~\ref{fig:unit_cells} ): a zig-zag and an armchair-like 
CNS~\cite{braga}. The Brillouin zone has been sampled with a 
converged grid of up to $1 \times 1 \times 12$ {\em k}-points. We 
have relaxed both the cell lattice vectors and the atomic positions, 
thus accounting for both the axial and radial response upon charge 
injection. We have also carried out a set of calculations where the 
lattice parameter was kept frozen, in order to explore complementary
experimental situations, free standing actuator (like in 
Ref.~\onlinecite{baughman}) vs. sensor constrained between two 
electrodes.


In agreement with the molecular dynamics results of Ref.~\onlinecite{braga}, 
we have chosen starting geometries with an internal radius of $\sim 20$\AA.
Larger systems, with a higher number of revolutions around the scroll
axis -~{\em i.e.} obtained wrapping a wider graphite sheet~- cannot be
efficiently handled within DFT. However, the fundamental mechanisms and 
driving forces associated with the geometrical expansion/contraction 
upon charge injection are already qualitatively captured by the systems 
studied. Nonetheless, we have chosen CNS geometries which already present 
the critical overlap between sheet layers that assures the scroll 
formation~\cite{braga}. We have injected a net charge of up to 
$\pm$ 0.055 $\mid$ $e$ $\mid / \text{ atom}$ into the systems by 
adding/removing electrons. In order to accommodate these extra/missing 
charges the carbon-carbon bond-lengths need to adjust their values and 
this is the origin of the electron-actuation phenomena.

The electro-actuation response relies on the competition between two
different effects: an {\em electronic} actuation, driven by the 
depletion/population of bonding/anti-bonding states, and a purely
{\em electrostatic} actuation, which originates from Coulomb repulsion. 
In the low-injection regime a contraction is expected upon electron removal
(hole injection) due to the lack of bond completion, {\em i.e.} creation
of dangling bonds, which makes the region locally more reactive.
Injecting electrons, on the other hand, results in a more complex situation. 
Adding electrons in anti-bonding states generates an electronic repulsion
and some bond-lengths will elongate, but this, in turn, could stretch 
some of the neighboring bonds. In general, for sp$^2$ carbon based structures 
the best compromise is alternating short and long bonds, even 
at the cost of breaking higher symmetries. This behavior is well-known 
for conducting polymers; for instance, in polymers containing benzenoid 
rings charge injection transform them into quinoid structures (with a 
well pronounced alternate of short and long bonds)~\cite{galvao}. Besides 
that, the picture is further complicated by the interplay
with Coulomb forces induced by the extra charge injected that will
tend to expand the scroll by pushing apart the overlapped layers.
Full quantum calculations are then necessary to quantify these effects.

\begin{figure}[t]
\begin{center}
\epsfxsize=14cm
\epsffile{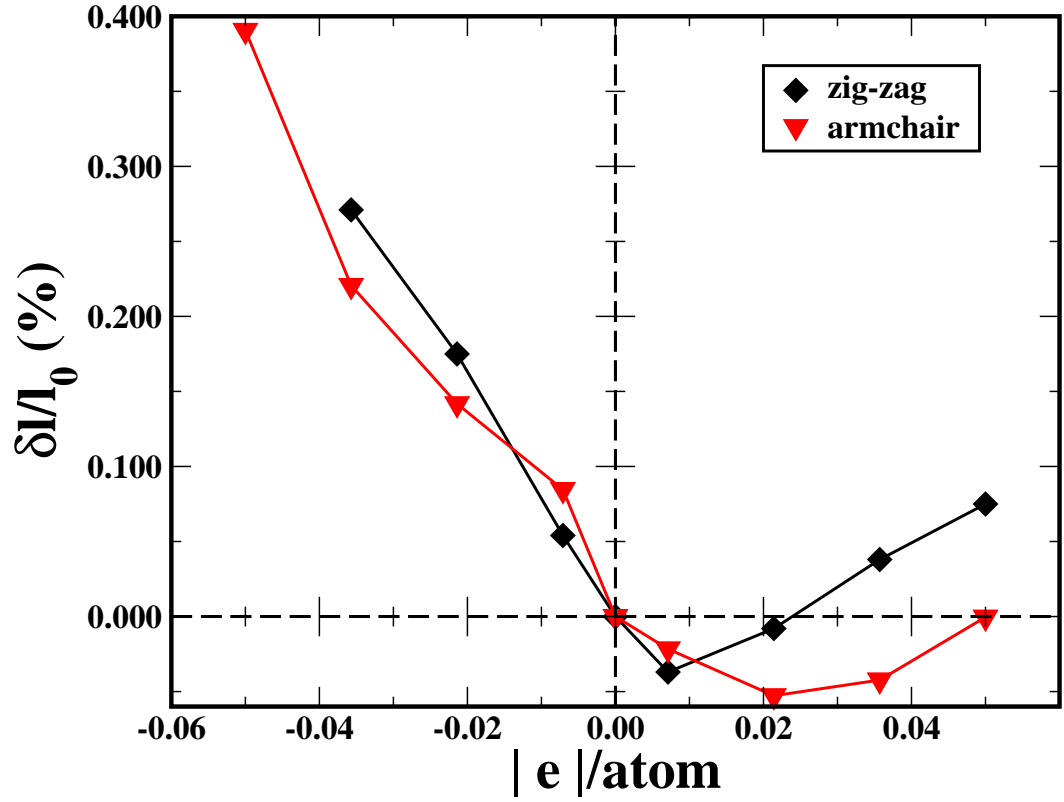}
\end{center}
\caption{(Color online) Relative variation of the unit cell lattice
         parameter $l$ upon charge injection with respect to its
         equilibrium value in the neutral state $l_0$.}
\label{fig:axis}
\end{figure}

In Fig.~\ref{fig:axis} we present the results for the relative axial 
variation of the lattice parameter ($\delta l/ l_0$). The dependence on 
the injected charges follows the general trends exhibit by (5,5) and 
(12,0) CNTs reported in Ref.~\onlinecite{verissimo}. CNSs expand to 
accommodate the extra electrons and slightly contract when holes are 
injected. In the high-injection regime the Coulombian repulsion dominates 
and the CNSs expand regardless of the sign of the injected charge. Most 
importantly, the magnitude of response is very similar to the case of 
CNTs and can reach values of about 0.2~-~0.3~\%.

\begin{figure}[t]
\begin{center}
\epsfxsize=14cm
\epsffile{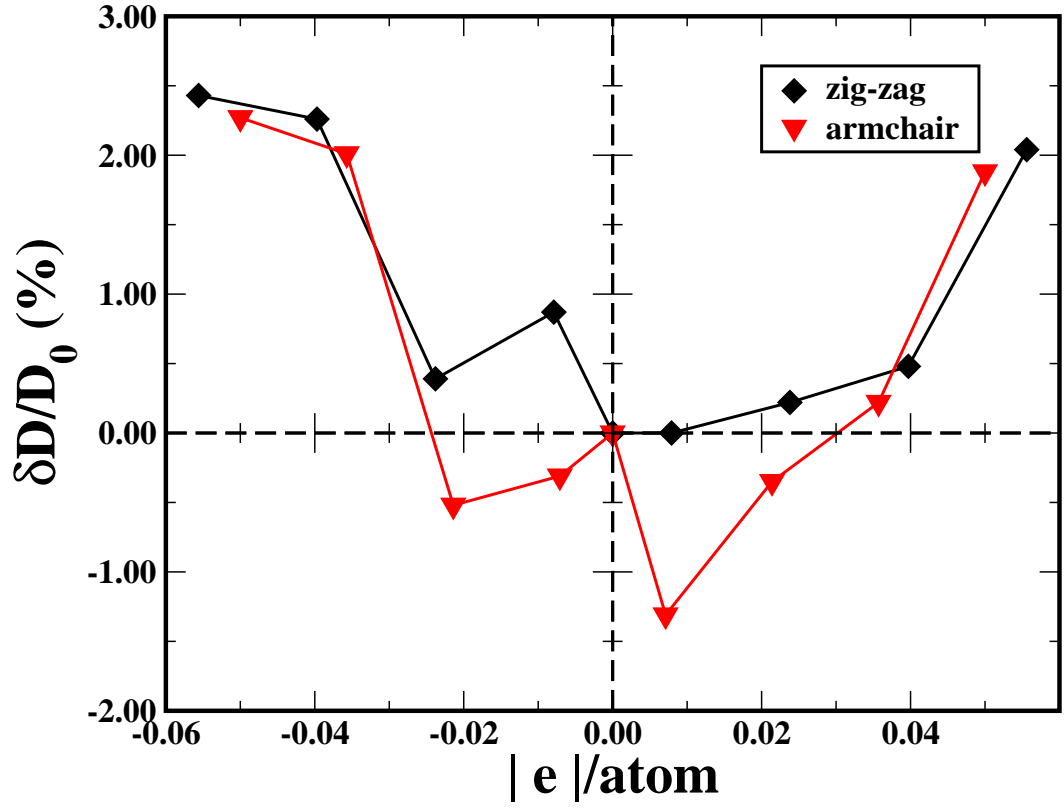}
\end{center}
\caption{(Color online) Relative variation of the CNS diameter
         parameter $D$ upon charge injection with respect to its
         equilibrium value in the neutral state $D_0$. The
         diameter $D$ has been defined as the maximum distance
         between two carbon atoms with same axial coordinate.}
\label{fig:diam}
\end{figure}

On the other hand, for what concerns the radial response (which is shown 
in Fig.~\ref{fig:diam}), the behavior of CNSs is completely different from 
CNTs. While for CNTs the actuation response is almost equally distributed 
between the axial and radial parts, we have found the latter to be up to 
one order of magnitude more intense for CNSs~\cite{cnt}, being approximately 
2.5~\% for the highest injected charge considered.
As can be seen, for small values of charge injection
armchair and zig-zag CNSs behave differently, one contracts while the other 
expands, thus the different topologies are still playing an important role. 
For higher charge values -~where the interlayer Coulomb interactions are 
expected to dominate~- both structures converge to almost the same values, 
recovering an almost linear behavior; again a close parallel with the 
behavior of doping conducting polymer is observed \cite{handbook}.

The data reported in Fig.~\ref{fig:diam} correspond to calculations 
where the lattice parameter was constrained to its 
equilibrium value in the neutral state. This arrangement is intended 
to mimic the situation where the CNS has its extremities clamped,
{\em i.e.} a suspended scroll. On the other hand, relaxing the lattice
better approaches the situation where the CNS is on a surface and can 
freely move in the axial direction too. In this case, however, the 
dependence of the diameter on the injected charge turned out to much 
more irregular, especially in high-charge regime, and it is often 
accompained by an {\em elliptization} of the structure. The radial 
response is sometimes even larger, but difficult to reliably associate 
with the injected charge. In other words, free-standing CNS on a surface 
behave axially as CNTs and have an enhanced, but noisy radial sensitivity; 
suspended CNSs, axially constrained, exhibit a giant and ordered radial 
electro-mechanical response.

\begin{figure}[t]
\begin{center}
\epsfxsize=14cm
\epsffile{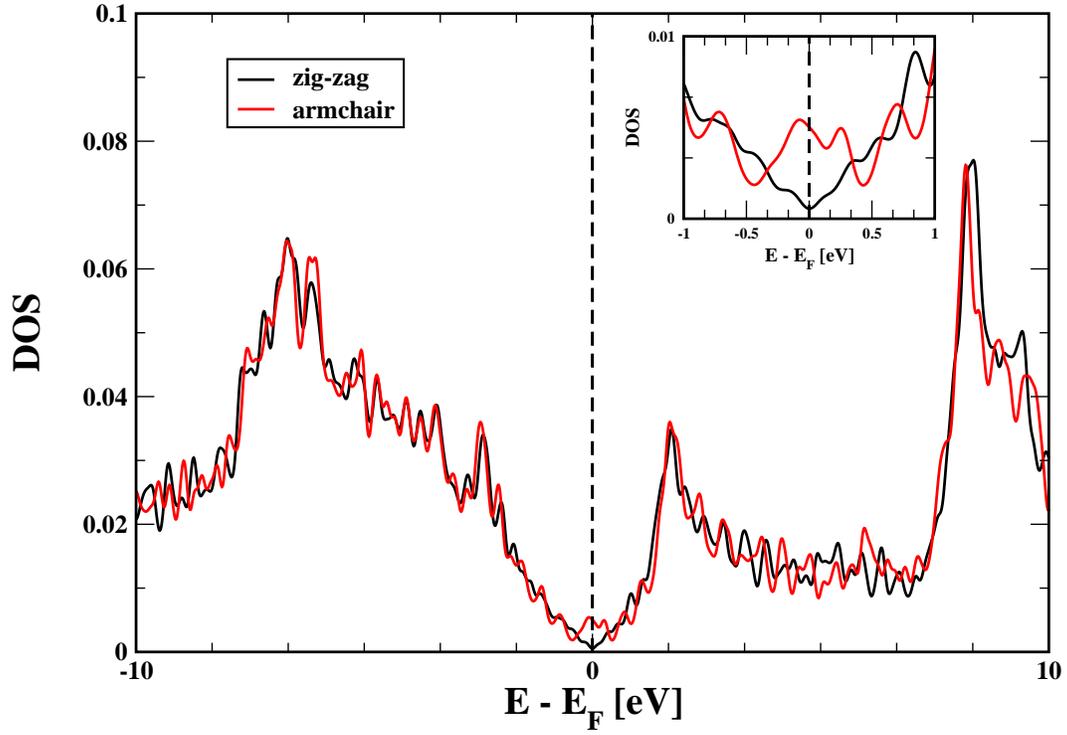}
\end{center}
\caption{(Color online) Electronic density of states of the armchair
         (red line)and the zig-zag (black line) CNS studied. The inset
         panel shows a magnified view around the Fermi level.}
\label{fig:dos}
\end{figure}

Recent DFT calculations of neutral scroll have attributed a metallic 
character to armchair CNSs~\cite{pan,e_struct_zig}. 
Our DFT calculations are in very good agreement with the results 
of Pan {\em et al.}~\cite{pan}, as shown in Fig.~\ref{fig:dos}. 
However, the scroll geometry that we have used differs significantly from
theirs (as we had to perform several full relaxations -~lattice and atomic
positions~- we had to use smaller structures). Hence, this is a hint that, 
at least at this scale, the electronic structure is at a first approximation 
insensitive to the size of the scroll. Zig-zag CNSs are predicted to have 
a small band-gap~\cite{e_struct_zig}. For both topologies, we have found 
that the electronic structure of CNSs around the Fermi level is determined 
by the border states (in agreement with the results reported by Pan {\em et 
al.}~\cite{pan} for what concerns armchair CNSs), in analogy with carbon 
nanoribbons from which they are derived by wrapping~\cite{pan,fujita,nakada}.

\begin{figure}[t]
\begin{center}
\epsfxsize=14cm
\epsffile{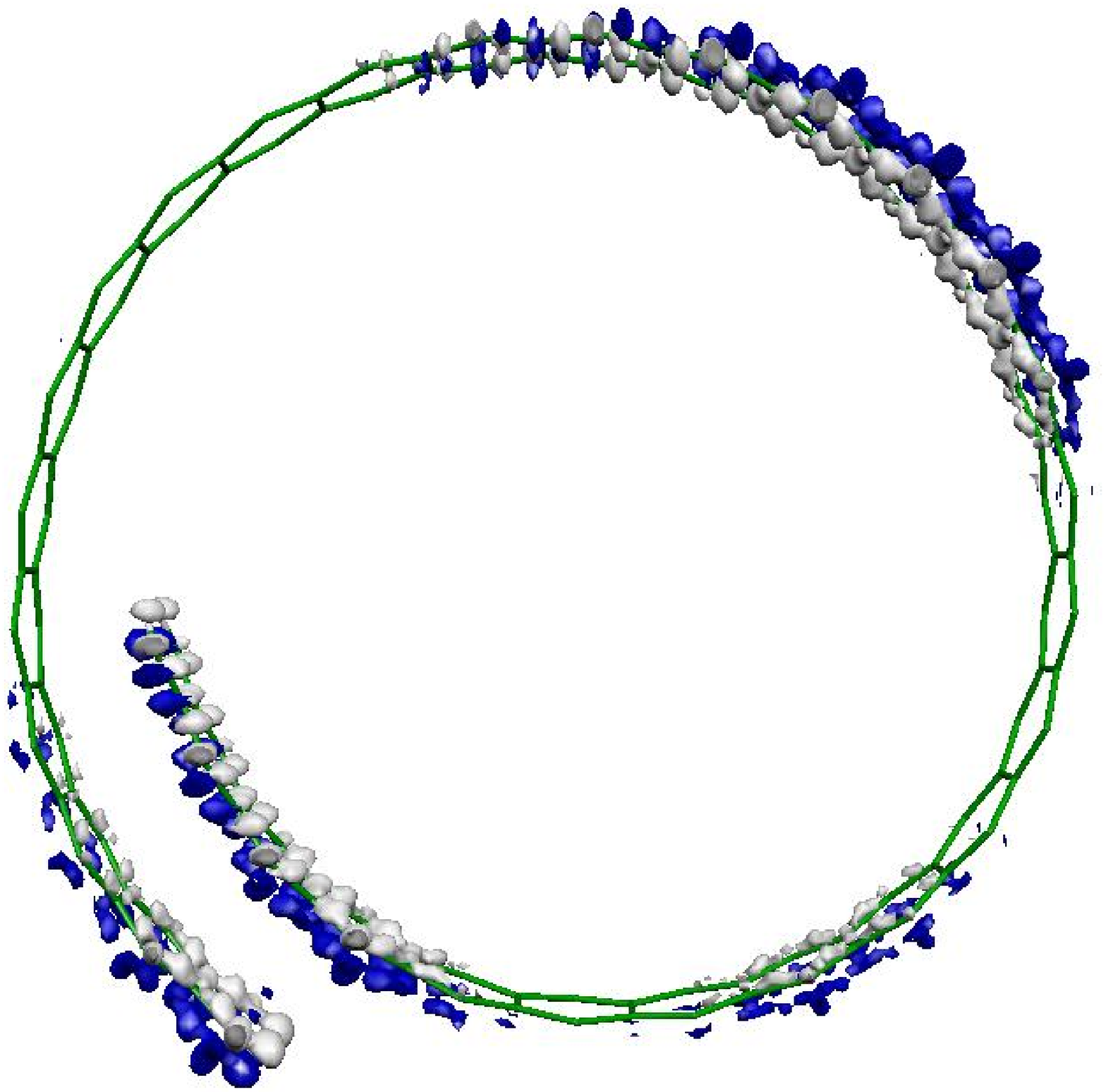}
\end{center}
\caption{(Color online) Distribution of the extra charge of -0.055
         $\mid e \mid/\text{atom}$ in a zig-zag CNS.}
\label{fig:diff_char}
\end{figure}

In order to gain a further insight on the electro-actuation process we have
studied the localization of the extra charge in the high-injection regime.
In Fig.~\ref{fig:diff_char} we have plotted the difference between the 
electronic density of a zig-zag CNS with 0.055 extra electrons per atom and 
the electronic density of the same CNS in the neutral state, {\em i.e.} the
extra electrons of the charged system. It can be 
seen that the excess charge accumulates in central region of the CNS and, 
with a clear discontinuity, close to the borders. On the other hand, 
analyzing the relative elongation of each individual bond we have found 
that almost all the bonds elongate in a similar way, despite the excess 
charge is not homogeneously distributed. The charge accumulated at the 
CNS boundary carries the large radial electro-mechanical response, due to 
efficient inter-layer Coulombian repulsion; the excess charge concentrated 
in the central region, on the other hand will only be responsible of
minor bond elongations, contributing to both axial and radial response. 
However, in the high-injection regime, the bond deformation is uniformly
distributed along the CNS. This is no longer true for low-injected
charge, where the predicted alternation of shortened and elongated bonds
is recovered. 
 
Therefore, the electro-mechanical actuation in the high-charge regime 
also originates in the charge accumulation in the central region of the 
CNS. In the studied geometry, presenting a limited layer overlap, this 
charge has only a minor impact in local bond elongation. 
In CNSs with a larger layer overlap also the extra charge
concentrated in the CNS center could contribute with an efficient
inter-layer Coulomb repulsion. Hence we do not discard that, in absence
of other side effects, {\em e.g.} inter-layer sliding, could not
lead to a larger electro-mechanical response.

In summary, we have carried out {\em ab initio} DFT calculations of the 
electro-actuation effect in carbon nanoscrolls. While the axial 
sensitivity of CNSs has very similar features than in carbon nanotubes, 
the radial response has a completely distinct behavior, reaching under
high-charge injection conditions the giant relative diameter variation 
of 2.0-2.5~\%, which is one order of magnitude higher than the values 
reported for CNTs. In suspended CNS is much easier to correlate the
charge injection with the electro-mechanical response, as the circular
symmetry of the scroll is qualitatively maintained for a wide range
of injected charge. Free-standing CNS are more naturally used as axial
actuators, even though in such a case their response is similar to
the case of CNTs. These results suggest that CNSs provide a simple and
flexible path towards the development of efficient electro-mechanical 
actuators at the nanoscale. We hope the present study to stimulate 
further experimental work to test these predictions.

\begin{acknowledgments}
R.~R. acknowledges the financial support of the Generalitat de Catalunya,
through a {\sc Nanotec} grant. V.R.C. and D.S.G. wish to thank
Dr. R. Giro and Prof. R.H. Baughman for helpful discussions. Work
supported in part by FAPESP, CAPES, CNPq, IN/MCT, and IMMP/MCT. 
\end{acknowledgments}

\newpage

\clearpage

\clearpage

\clearpage

\clearpage

\clearpage

\begin{thebibliography}{99}

\bibitem{ijima} S.~Ijima,
                Nature (London) {\bf 354}, 56 (1991). 

\bibitem{deheer}  W.~A.~de Heer, P.~Poncharal, C.~Berger, J.~Gezo, 
                  Z.~M.~Song, J.~Bettini, and D.~Ugarte,
                  Science {\bf 307}, 907 (2005).
	     
\bibitem{bacon} R.~Bacon,
                J. Appl. Phys. {\bf 31}, 283 (1960).
	     
\bibitem{zhou} O.~Zhou, R.~M.~Flemming, D.~W.~Murphy, C.~H.~Chen, 
               R.~C.~Haddon, A.~P.~Ramirez, and S.~H. ~Clanum,
               Science {\bf 263}, 1744 (1994).	     
	          
\bibitem{ruland} W.~Ruland, A.~K.~Schapper, H.~Hou, and A.~Greiner,
                 Carbon {\bf 41}, 423 (2003).
	     
\bibitem{shioyama} H.~Shioyama and T.~Akita,
                   Carbon {\bf 41}, 179 (2003).	     
	     	          
\bibitem{viculis} L.~M.~Viculis, J.~J.~Mack, and R.~B.~Kaner,
                  Science {\bf 299}, 1361 (2003).
	     
\bibitem{khang} Z.~Khang, E.~Wang, L.~Gao, S.~Lian, M.~Jiang, 
                C.~Hu, and L.~Xu
                J. Am. Chem. Soc. {\bf 125}, 13652 (2003).

\bibitem{lauin} J.~C.~Lauin, S.~Subnamoney, R.~S.~Ruoff, S.~Berber, 
                and D.~Tom\'{a}nek,
                Carbon {\bf 40}, 1123 (2002).

\bibitem{ameclinckx} A.~Ameclinckx, D.~Bennates, X.~B.~Zhang, 
                     G.~van~Thendeloo, and J.~van~Landuyt,
                     Science {\bf 267}, 1334 (1995).	     

\bibitem{mordkovich} V.~Z.~Mordkovich, M. Baxendale, S.~Yoshimura, 
                     R.~P.~H.~Chang,
                     Carbon {\bf 34}, 1301 (1996).	     
	     	          	     	     
\bibitem{maniwa} Y.~Maniwa, R.~Fujiwara1, H.~Kira1, H.~Tou1, E.~Nishibori, 
                 M.~Takata, M.~Sakata, A.~Fujiwara, X.~Zhao, S.~Iijima, 
                 and Y.~Ando,
                 Phys. Rev. B {\bf 64}, 073105 (2001).	     	     	          	     

\bibitem{tomanek} D.~Tom\'{a}nek, W.~Zhong, and E.~Krastev,
                  Phys. Rev. B {\bf 48}, 15461 (1993).

\bibitem{tomanek2} D.~Tom\'{a}nek, 
                   Physica B {\bf 323}, 86 (2002).

\bibitem{braga} S.~F.~Braga, V.~R.~Coluci, S.~B.~Legoas, R.~Giro, 
                D.~S.~Galv\~{a}o, and R.~H.~Baughman,
                Nano Lett. {\bf 4}, 881 (2004).

\bibitem{pan} H.~Pan, Y.~Feng, and J.~Lin,
              Phys. Rev. B {\bf 72}, 085415 (2005).

\bibitem{verissimo} M.~Verissimo-Alves, B.~Koiller, H.~Chacham, and
                    R.~B.~Capaz,
                    Phys. Rev. B {\bf 67}, 161401(R) (2003).

\bibitem{gartstein} Y.~N.~Gartstein, A.~A.~Zakhidov, and R.~H.~Baughman,
                    Phys. Rev. Lett. {\bf 89}, 045503 (2002).

\bibitem{baughman} R.~H.~Baughman, C.~X.~Cui, A.~A.~Zakhidov, Z.~Iqbal, 
                   J.~N.~Barisci, G.~M.~Spinks, G.~G.~Wallace, A.~Mazzoldi, 
                   D.~De Rossi, A.~G.~Rinzler, O.~Jaschinski, S.~Roth, 
                   and M.~Kertesz,
                   Science {\bf 284}, 1340 (1999).

\bibitem{siesta} P.~Ordej\'{o}n, E.~Artacho and J.~M.~Soler,
                 Phys. Rev. B {\bf 53}, R10441, (1996);
                 J.~Soler, E.~Artacho, J.~D.~Gale, A.~Garc\'{\i}a,
                 J.~Junquera, P.~Ordej\'{o}n and D.~S\'{a}nchez-Portal,
                 J. Phys.:~Condens. Matter, {\bf 14}, 2745 (2002);
                 see also http://www.uam.es/siesta/

\bibitem{troullier:martins} N.~Troullier and J.~L.~Martins,
                            Phys. Rev. B {\bf 43}, 1993 (1991).
     
\bibitem{galvao} D. S. ~Galv\~ao, D. A. ~dos Santos, B. ~Laks, 
                 C. P. ~de Melo, M. J. ~Caldas,
                 Phys. Rev. Lett. {\bf 63}, 786 (1989). 

\bibitem{cnt} A detailed comparison with equivalent results for
              CNTs is not possible. In Ref.~\onlinecite{gartstein}
              the authors restrict to very small charge injection
              ($\mid e \mid/\text{atom} = 0.005$). Ref.~\onlinecite{verissimo}
              investigates a wider range of electron/hole injection,
              comparable to ours, but they only report the results
              concerning the axial response, suggesting, however,
              that the radial behavior is qualitatively similar.

\bibitem{handbook} {\em Handbook of Conducting Polymers}, 
                   edited by T. A. Skotheim (Dekker, New York, 1986). 

\bibitem{e_struct_zig} No data other than ours on the {\em ab initio} 
                       electronic structure of zig-zag CNSs are available
                       at present. We are currently performing an
                       extensive study of a wide range of scroll geometries
                       to generalize our conclusions. 

\bibitem{fujita} M.~Fujita, K.~Wakabayashi, K.~Nakada, and K.~Kusakabe,
                 J. Phys. Soc. Jpn. {\bf 65}, 1920 (1996).

\bibitem{nakada} K.~Nakada, M.~Fujita, G.~Dresselhaus, and M.~S.~Dresselhaus,
                 Phys. Rev. B {\bf 54}, 17954 (1996).

     
\end{thebibliography}
\end{document}